\documentclass[english]{emulateapj}
\usepackage{color}
\usepackage{amssymb}
\usepackage{graphicx}

\makeatletter

\providecommand{\tabularnewline}{\\}

\usepackage{babel}

\usepackage{babel}

\usepackage{babel}

\makeatother

\usepackage{babel}
\begin{document}

\title{Age and mass segregation of multiple stellar populations in galactic
nuclei and their observational signatures }

\author{Hagai B. Perets and Alessandra Mastrobuono-Battisti }
\begin{abstract}
Nuclear stellar cluster (NSCs) are known to exist around massive black
holes (MBHs) in galactic nuclei. They are thought to have formed through
in-situ star formation following gas inflow to the nucleus of the
galaxy and/or through the infall of multiple stellar clusters. Here
we study the latter, and explore the composite structure of the NSC,
and its relation to the various stellar populations originating from
its progenitor infalling clusters. We use N-body simulations of clusters
infall, and show that this scenario may produce observational signatures
in the form of age segregation: the distribution of the stellar properties
(e.g. stellar age and/or metallicity) in the NSCs reflect the infall
history of the different clusters. The stellar populations of clusters
infalling at different times (dynamical ages), are differentially
segregated in the NSC, and are not fully mixed even after few Gyrs
of evolution. Moreover, the radial properties of stellar populations
in the progenitor cluster are mapped to their radial distribution
in the final NSC, potentially leading to efficient mass segregation
in NSCs, even those where relaxation times are longer than a Hubble
time. Finally, the overall structures of the stellar populations present
non-spherical configurations and show significant cluster to cluster
population differences. 
\end{abstract}

\section{Introduction}

Nuclear stellar clusters (NSCs) hosting massive black holes (MBHs)
are thought to exist in a significant fraction of all galactic nuclei.
The build-up of such dense clusters is likely linked to the growth
of the MBH, and the evolution of the galaxy and its nucleus, as suggested
by statistical correlations between their properties (e.g. \citealp{fer+00,tre+02}).
Two scenarios were suggested for the origin of nuclear clusters: (1)
The cluster infall scenario, in which stellar clusters inspiral to
the galactic nucleus, disrupted, and thereby build up the nuclear
cluster (\citealp[bek+][]{tre+75,cap93,bek+04,aga+11,ant+12,ant+13,gne+13}
and references therein) the inspiraling clusters may also be the NSC
of two galaxies in a merger scenario). (2) The nuclear star formation
scenario, in which gas infalls into the nucleus and then transforms
into stars through star formation processes (\citealp{loo+82}; possibly
in a disk like configuration, e.g. observations of the Milky Way NSC;
\citealp{lev+03,lu+09,bar+10}). Over time the in-situ star formation
builds up the NSC. Naturally, both processes can work in concert,
and both could be important for the formation and evolution of NSCs.

In this paper we explore the cluster infall scenario by means of N-body
simulations, and neglect star formation processesand
their effects, which are beyond the scope of this paper. Previous
studies have dealt with the global structure and build-up of NSCs
from inspiral of globular clusters (GCs). Here we focus on the multiple
stellar populations in NSC and their mapping and relations to their
original host GCs. We study the evolution of the multiple GC populations
both during the evolution of the NSC and in its final form. In particular,
we show that the cluster infall scenario introduces population segregation
in NSCs, and provides signatures of the cluster infall history in
the radial distribution of the stars in the NSC. In addition we suggest
that the formation of NSCs from cluster infall can produce mass segregated
NSCs, even in cases where two-body relaxation processes are too slow.
We suggest that such variations in the kinematic properties of different
stellar populations might be observable and serve as fossilized evidence
for the evolution and build-up of galactic nuclei. 

In the following we begin by a brief description of our cluster infall
scenario (described in more details in Paper I). We then present the
distribution of the stellar populations of the different infalling
cluster in the fully formed NSC, and show the existence of the age-segregation
and mass segregation phenomena. Finally, we discuss our results, discuss
their implications and summarize.

\section{Simulation of the cluster infall formation of a nuclear cluster}

Our modeling of the cluster infall formation of an NSC follows the
same methods, and make use of the same code as used in paper I \citep{ant+12},
where detailed description of the initial conditions of GCs and the
galaxy model of the background stellar population can be found. In
brief, we used direct N-body simulations (using the $\phi$GRAPE code
\citealp{har+07}) of the consecutive infall and merging of a set
of 12 single-mass globular clusters each starting from a galactocentric
distance of $20$~pc. One simplification that we make is taking a
constant time interval between cluster infalls, as done in previous
works (e.g. Paper I). The total mass of these clusters sums to $\sim1.5\times10^{7}M_{\odot}$,
which is roughly the observed mass of the Milky Way nuclear star cluster
\citep{gen+10}. The mass of the GCs is comparable to currently observed
Milky Way GCs \citep[see][ for Milky-Way GC paramters]{gne+97}, though
a better comparison would be to young superclusters (or their leftover
nuclei) that would have inspiarlled to the nucleus (e.g. \citealp{kro98}),
and are currently observed only in other galaxies (e.g. young super
star-clusters such as R136 observed in the LMC). with a MBH of $4\times10^{6}$${\rm M_{\odot}}$
. After the first cluster had spiraled in to the center, we let the
system reach a nearly steady state (as evaluated via Lagrange radii),
and then added a second cluster. We iterated this procedure until
all clusters accumulated and merged to form an NSC around the central
MBH. We note a few differences from the original simulations discussed
in paper I: (1) The scaling of the relaxation time used in paper I
assumed a fully collisional simulation, where as in effect a softening
radius of 0.01 pc is used; we now scale the relaxation times correctly,
accounting for the lower cut-off of the Coloumb logarithm due to the
softening radius used. The times used in paper I were therefore 2.9
times shorter than the now corrected estimates. (2) We now assume
a random distribution of the initial inclinations and phases of the
inspiraling clusters, rather than a the contrived model in paper I;
the initial conditions are shown in Table \ref{tab:Initial-orbital-paramters}.
(3) The total scaled time of the simulation is 12 Gyrs, comparable
to the age of the Galaxy; the time passed since the infall of the
last cluster until the end of the simulation is shorter than the relaxation
time of the NSC, but we have also followed the simulation up to the
relaxation time of the NSC. We emphasize
that the times used here are only an approximation, based on scaling
arguments of the relaxation time, and should not be treated as an
accurate time representation.

We have made two realizations of the infall scenario for the formation
of the NSC. Both realizations show generally similar results, and
we present results only from one of them (see Table \ref{tab:Initial-orbital-paramters}
for the relevant initial conditions). We briefly remark on some differences
observed between the two realizations. 

\begin{table}[b!]
\begin{tabular}{cccc}
\hline 
n  & $\Omega$ (deg) & $i$ (deg) & $R_{t}$(pc)\tabularnewline
\hline 
1  & 82.4  & 60.7 & 1.29\tabularnewline
2  & 327.7  & 178.7 & 1.29\tabularnewline
3  & 76.2  & 139.5 & 1.29\tabularnewline
4  & 290.6  & 171.3 & 1.34\tabularnewline
5  & 335.4  & 24.6 & 1.48\tabularnewline
6  & 300.6  & 18.2 & 1.54\tabularnewline
7  & 343.9  & 173.9 & 1.55\tabularnewline
8  & 47.9  & 128.9 & 1.6\tabularnewline
9  & 272.0  & 2.3 & 1.78\tabularnewline
10  & 41.3  & 139.0 & 1.80\tabularnewline
11  & 300.9  & 153.5 & 1.85\tabularnewline
12  & 318.2  & 120.2 & 1.86\tabularnewline
\hline 
\end{tabular} \caption{\label{tab:Initial-orbital-paramters}Initial orbital parameters of
the infalling clusters (inclination i and longitude of ascending node
$\Omega$), and their tidal disruption radius
as found from the simulations.}
\end{table}

\section{Results}

In this study we explore the mapping between the properties of stellar
populations in NSC, and their relation to the initial characteristics
of the stellar populations in the progenitor clusters. We focus on
two points in time during the NSC evolution; the first after the infall
and initial relaxation of the last cluster, and the second, after
more relaxation has occurred at an age comparable to a Hubble time.
Our findings show that the dynamical history of the cluster infall
is still reflected in the radial distribution of the stellar populations
in the NSC, even Gyrs after the last infall. 

\begin{figure}
\includegraphics[scale=0.4]{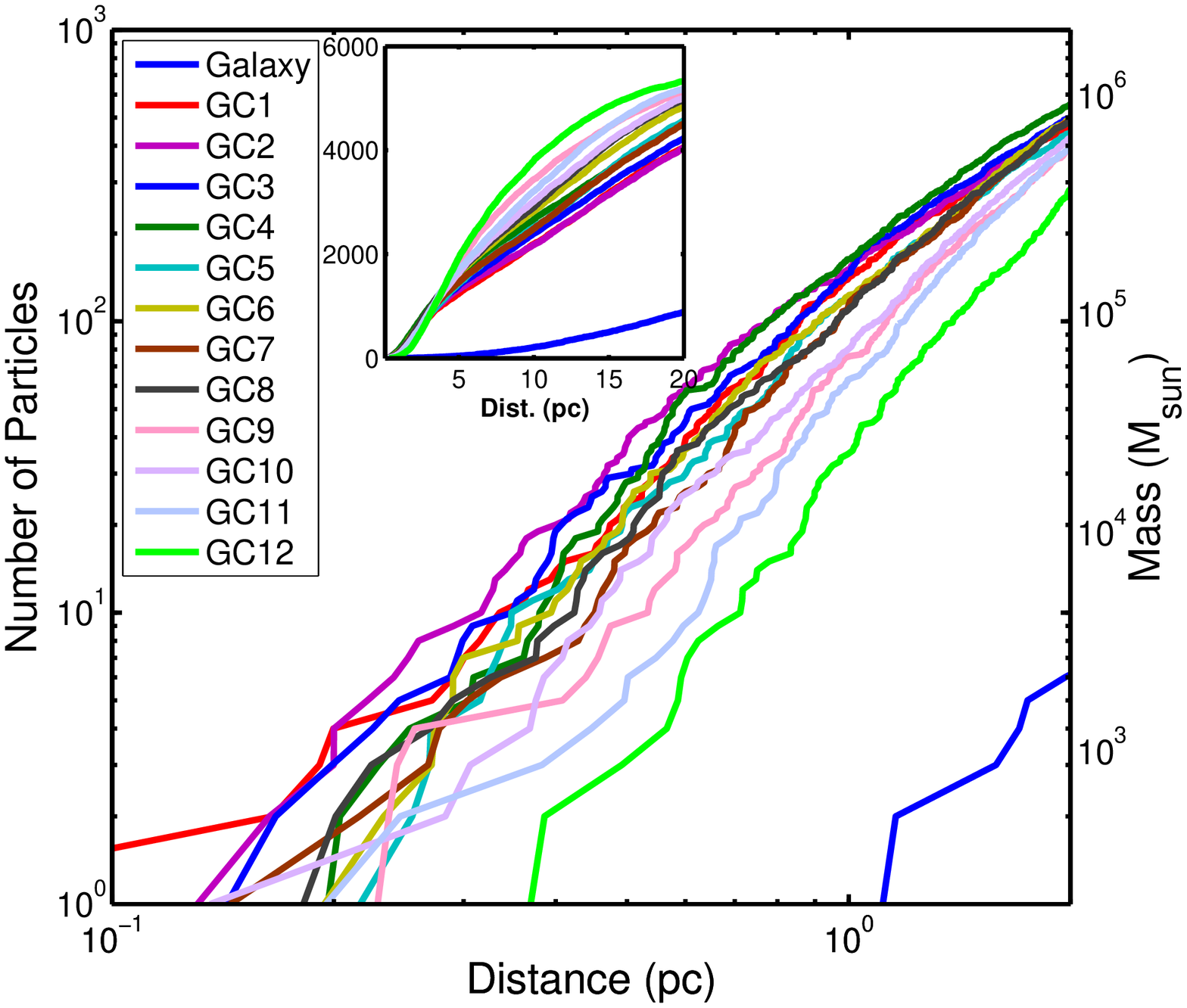}

\includegraphics[scale=0.4]{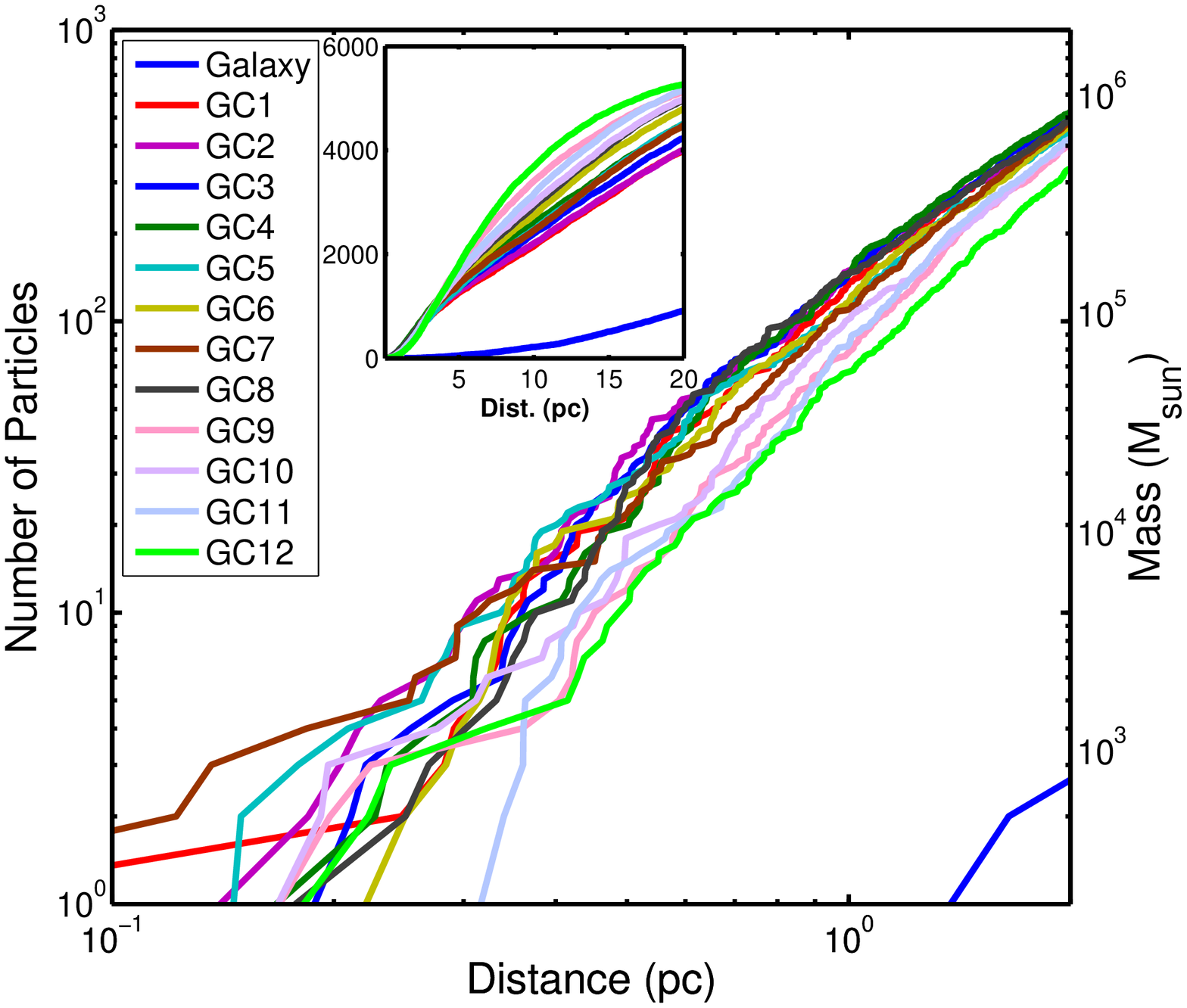}\caption{\label{fig:Radial-distribution}Radial distribution of the infalling
clusters stellar populations. Top: the radial distribution following
the last cluster infall (after the initial relaxation). Bottom: The
same at 12 Gyrs. As can be seen in both cases the clusters populations
show clear differences, with the earlier infalling clusters showing
systematically more compact configuration in the central region ($<2$
pc), and then an opposite behavior outside. Insets
show the large scale (20 pc) distribution; note linear scales in insets.}
\end{figure}

\begin{figure}
\includegraphics[scale=0.4]{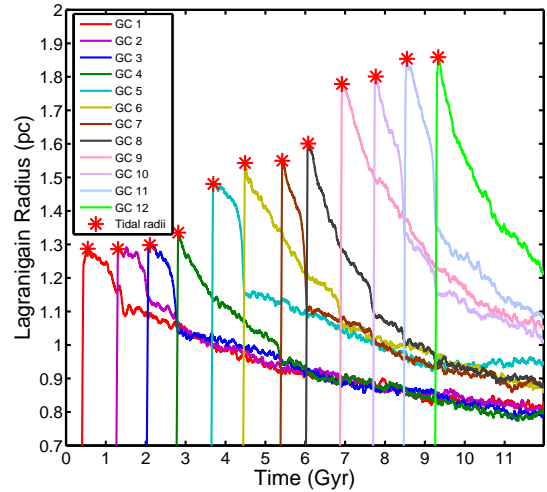}\caption{\label{fig:Lagrange}
Evolution of the Lagrangian radii for the central
hundred stars of each of the clusters.The
initial Lagrange radius is 0.2 pc for each cluster. Once the cluster
is disrupted (see star symbols), the Lagrange radius transitions from
being relative to the location of the central density of the cluster
to effectively become the distance from the MBH, hence the rise from
0.2 pc (no seen) to the tidal radius central density of each cluster.
As can be clearly seen, the stellar population of the earlier falling
clusters is progressively more centrally concentrated than the populations
of later falling clusters. }
\end{figure}

We find that stars in the NSC originating from early-infall clusters
are more segregated to the center of the NSC than their stellar counterparts
from late-infall clusters. This is true both after the last infall
and even later after a few Gyrs of evolution (See Fig. \ref{fig:Radial-distribution}).
In particular the cluster inner population (central hundred particles),
show a clear segregation when comparing early and late infalls, with
relatively little additional mixing over the last two Gyrs of evolution
(See the evolution of the Lagrangian radii of these populations in
the NSC in Fig. \ref{fig:Lagrange}). Note that the age segregation
is reversed between the inner regions of the NSC and the outer regions,
due to the later (earlier) stripping of the early (late)-infalling
clusters. 

As can be seen in Fig. \ref{fig:Lagrange},
the infall of each new cluster affects the evolution of the stellar
population of the previous infalling cluster, effectively ``compressing''
it into a more compact configuration around the MBH. The last infalling
cluster did not experience such a later infall, and the distribution
of its stars is significantly less segregated. Observing such a distinct
population in a galactic nucleus could therefore provide an interesting
clue on a relatively recent infall. The more robust results apply
for the earlier 11 clusters, in which the age segregation signatures
can be observed even long after their infall. 

\begin{figure}
\includegraphics[scale=0.4]{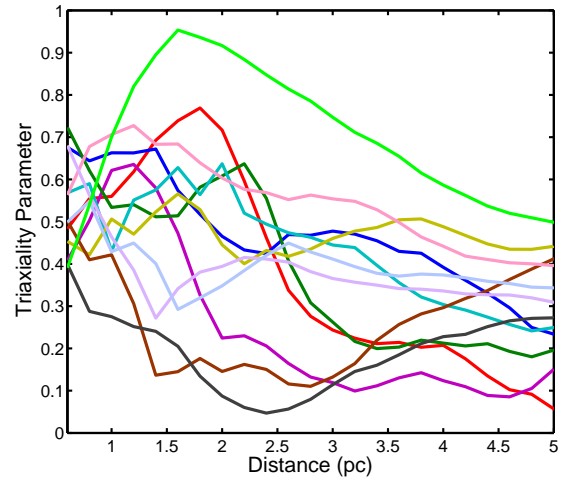}\caption{\label{fig:triaxial}Triaxiality of the NSC stellar populations. The
triaxiality parameter of the different cluster populations (smoothed
for clarity with a 1 pc window) is shown as a function of distance
from the MBH. As can be seen the different populations show distinctly
different behavior, and significant triaxiality. In particular, some
of the cluster show highly anisotropic properties; somewhat resembling
thick-disk-like structures in some cases even Gyrs after their infall. }
 
\end{figure}

We also find that the 3D structure of the stellar populations of each
of the GCs could significantly differ. In Fig. \ref{fig:triaxial}
we show the triaxiality parameter (see Paper I) for each of the GC
populations as a function of the distance from the MBH. As can be
seen, such structure could vary significantly even between consecutive
infalling clusters. 

We note that some bunching of the population of several consecutive
clusters can be observed (e.g. GCs 1-4, then GCs 5-8, and then 9-11).
This bunching relates to the specific initial conditions for each
cluster (i.e. its inclination and orbital phase; i and $\Omega$),
and different bunching is observed when different random initial conditions
are used (not shown). However, this interesting phenomena do not affect
the overall age and mass segregation processes discussed here, and
are beyond the scope of this paper. 

\section{Discussion}

\subsection{Age segregation}

An additional aspect of NSC build-up from cluster infall is the mapping
of the infalling clusters structure to the final configuration of
the NSC. Typically stars at a distance $R_{c}$ (which contains $M(<R_{c})$
of the cluster mass) from the center of their infalling host cluster
are stripped by the tidal forces due to the combined mass of the MBH
and the stellar mass of the NSC, $M_{NSC}$ at the stripping radius,
$R_{s}$, defined as the tidal radius for stars at that position:

\begin{center}
\begin{equation}
R_{s}=\left(\frac{M_{BH}+M_{NSC}(<R_{s})}{M_{c}(<R_{c})}\right)^{1/3}R_{c}.\label{eq:Rt}
\end{equation}

\par\end{center}

At early stages, the NSC mass is dominated by the MBH ($M_{BH}\gg M(<R$))
and the early falling clusters should have very similar tidal radii.
At later stages the NSC is slowly built up from the infalling clusters,
until it reaches a mass comparable and even larger than the MBH ($M_{NSC}\sim2-3\times M_{BH}$).
We therefore expect stars from later clusters to be stripped earlier,
at a stripping radius which is up to $2^{1/3}-3^{1/3}$ (1.26-1.44)
larger than the corresponding stripping radius for stars from earlier
GCs (i.e. initially at comparable distances from the GC center in
both clusters). This expectation is consistent with our results in
table 1 and Fig. \ref{fig:Lagrange}; indeed
the first infalling clusters show the same tidal radii, and the ratio
between the tidal radii of the last and the first infalling clusters
is $1.86/1.29\simeq3^{1/3}$.
This phenomenon produces a dynamical age gradient, where\textbf{ stars
from earlier infalling clusters are stripped at later stages, and
dominate inner regions of the NSC}. If earlier falling clusters are
older (e.g. they formed earlier and therefore spiraled earlier to
the nucleus) then such gradient would be translated into a stellar
age gradient. Alternatively, if earlier falling clusters differ in
metallicity/color than their later counterparts (e.g. later falling
cluster arrive from larger distances where colors differ; e.g. \citealp{bal+94}
), such gradient would be translated into a color/metallicity gradient
in the NSC. The dynamical age segregation process may therefore give
rise to potentially observable stellar age/metallicity gradients.

We do note an important complication with this simplified picture.
In particular, more compact and massive GCs will be stripped later
(see Eq. \ref{eq:Rt}), and may therefore dispose their stars closer
to the MBH, compared with the stars of the earlier, more dispersed
clusters. Note, however, that less compact clusters are less likely
to inspiral to the nucleus from large distances during a Hubble time
(see \citealp{ant+12}), and therefore we expect only the most dense
massive clusters to contribute to the stellar population in the NSC.
Moreover, given the shorter inspiral time of more massive clusters,
these are likely to arrive earlier, on average, than their lighter
counterparts, which would also contribute to their later stripping.

\subsection{Mass segregation}

Irrespective of the age segregation process discussed above, the mapping
between the structure of the infalling GC and the structure of the
stripped population of stars in the NSC suggested by Eq. \ref{eq:Rt}
could have an important role in building mass segregated NSCs. It
is thought that massive clusters are likely to be mass segregated
either primordially or from an early stage in their evolution \citealp[e.g. ][and references therein]{bau+08,all+09,por+10};
and in even in the absence of such early mass segregation, GCs may
have sufficient time to segregate during their infall. Such mass segregation
could potentially be mapped into a mass segregation of galactic nuclei
stellar population. Indeed, the locations of stars initially outside
the GC cores are strongly correlated with their positions in the galaxy
nucleus at the end of the simulation (linear correlation coefficients,
$R$, almost monotonically rise between the first infalling clusters,
$R=0.48$ to the last infalling cluster, $R=0.18)$. Note, however,
that we find such direct mapping of closer in stars do not hold for
stars initially inside the GC cores. These stars would still be more
centrally concentrated around the MBH at the end of the simulation
than stars initially residing in the outer parts of the GCs (by a
factor of 2-3), but among these core stars any direct correlation
between their initial position in the GC and their final position
in the galaxy nucleus is lost. 

The infall scenario would therefore suggest that \textbf{NSCs are
likely to be mass segregated even when the relaxation times in such
NSCs are longer than a Hubble time} (true for most NSCs hosting MBHs
with $M_{BH}>{\rm few}\times10^{7}$). More generally, NSCs could
be more mass segregated than expected from typical relaxation processes
(e.g. \citealp{bah+77}; though strong mass segregation process may
also contribute; \citealp{ale+09}). As a side note, it is interesting
to point out that clues for an extreme mass segregation are apparent
in the stellar population of the Galactic NSC (see \citealp{ale07}
for a discussion). 

Such dynamical-mapping age segregation could have an important role
in leading to the concentration of massive stars, and in particular
stellar black holes in the central region of nuclear cusps around
MBHs. The built-up of a centrally concentrated dark cusp made of black
holes is therefore an interesting potential outcome of the cluster
infall scenario for NSC formation (see Antonini 2014, in prep. for
a detailed discussion of these issues). We also note that in cases
an infalling GC harbors an intermediate mass black hole ($10^{3}-10^{4}$${\rm M_{\odot}}$),
it could bring stars much closer in to the NSC MBH, thereby producing
an even more compact configuration (see Mastrobuono and Perets, in
prep. for such a scenario).

\subsection{Relaxation time vs. mixing time}

Following the cluster infall formation of NSC, they may continue to
evolve through two-body relaxation processes. In theory, such later
evolution may progressively erase some or all of the signatures of
the cluster infall scenario discussed above given sufficient time
for evolution. However, as shown above, we find that signatures of
the infall scenario are observed even at the age of the universe.
Moreover, even at later times, comparable to the relaxation time of
the NSC after the last infall, the NSC still shows clear signatures
of the dynamical-mapping age and mass segregation. This may appear
counter intuitive, as one might expect any initial conditions in the
cluster to be erased after a relaxation time. Relaxation time is defined
as the time it takes a star to change its kinetic energy by the order
of itself (e.g. \citealp{bin+87}). However, in order for a star in
the outer region of an NSC to be transported to the inner region of
the NSC, it needs a much more significant change in energy, i.e. of
the order of the energy of the star in the inner region, were the
gravitational potential and velocities are much higher. As a first
simplified approximation for the time it takes a stellar population
at some distance $r_{out}$ from MBH to mix with another stellar population
at an inner region, $r_{in},$ one can replace the velocity dispersion
in relaxation time formula at the position $r_{out}$ with that in
the position $r_{in}$ while keeping the number densities and the
relative velocities between stars, the same i.e. the ``mixing time''
for the two populations would be defined by 

\[
t_{mix(r_{out},r_{in})}=\left(\frac{\sigma(r_{in})^{2}}{\sigma(r_{out})^{2}}\right)t_{r_{out}}\thickapprox\left(\frac{r_{out}}{r_{in}}\right)t_{r_{out}},
\]
where $t_{r_{out}}$ is the relaxation time at $r_{out}$ and the
last equality is obtained for regions where the MBH dominates the
gravitational potential of the NSC (i.e. up to the MBH influence radius).
A more accurate definition would account for the changing diffusion
time as the stellar environment changes during the diffusive transport
of a star from one environment to another; a full discussion of the
mixing time is beyond this scope and will be discussed in details
elsewhere. Irrespective of the accurate definition, it is clear that
$t_{mix}>t_{r}$ ; stellar population segregation could therefore
survive much longer than a relaxation time. One should note, however,
that every infall of an additional cluster does not only bring new
stars to the NSC, that would slowly change the two-body relaxation
time, but can produce significant changes in the gravitational potential
on dynamical timescales. Though this may not significantly affect
the inner regions deep in the potential of the MBH+NSC, such changes
may give rise to a more violent relaxation in the outer regions that
can mix the stellar populations much more efficiently than two-body
relaxation processes. One would therefore expect segregated populations
to be more pronounced in the inner regions of NSCs, and around more
massive MBHs.

\subsection{Structure of the multiple stellar populations}

The particular final structure of the stellar population of each GC
is complex, as can be seen in Fig. \ref{fig:triaxial}, and its detailed
exploration is beyond the scope of this letter. Here we only note
that the significant differences between these structures could provide
an additional signature for the multi-cluster infall scenario, similar
to the radial segregation discussed above. In particular infalling
clusters can produce thick flattened structures with varied orientations,
possibly related to ``disky'' like structures are observed in galactic
nuclei and clusters (see \citealp{mas+13} for discussion of the evolution
of such disks).

\section{Summary}

In this \textit{letter }we explore the signatures and the implications
of the cluster infall scenario on the structure of nuclear stellar
clusters and their multiple stellar populations. We use N-body simulations
to study the infall of 12 globular clusters into a galactic nucleus
hosting a MBH of $4\times10^{6}$ ${\rm M}_{\odot}$, and we follow
the evolution of the stellar populations from each cluster and their
final distribution in the NSC. We find that the infall history is
reflected in the final structure of the NSC, where stellar populations
from earlier falling clusters are more concentrated in the central
parts of the NSC compared to late ones. This dynamical age segregation
process can potentially leave behind a signature in the form of an
age and/or metallicity radial gradient in the NSC stellar population.
The stellar population of each cluster forms a non-spherical complex
structure, which behavior significantly differs from one cluster population
to another. In addition, any primordial/early mass segregation in
the infalling GCs is mapped into a mass segregated populations in
the galactic nucleus; in particular even NSCs where relaxation time
is longer than a Hubble time could show a mass segregated stellar
population, which could not arise from two-body relaxation processes.

\acknowledgements{We would like to thank Fabio Antonini for for helpful discussions
and comments. We would also like to thank
the referee for helpful comments and suggestions that improved this
manuscript. We acknowledge support from the I-CORE Program of the
Planning and Budgeting Committee and The Israel Science Foundation
grant 1829/12. HBP is a Deloro and BIKURA fellow. AMB is partly supported
by the Lady Davies Foundation.}

\bibliographystyle{apj}


\end{document}